\documentclass[aps,prx,twocolumn,groupedaddress,amsmath,superscriptaddress,10pt,longbibliography]{revtex4-2}

\usepackage{graphicx}
\usepackage{amssymb,amsthm,mathtools,dsfont}
\usepackage{physics}
\usepackage{comment}
\usepackage{xcolor}
\usepackage[colorlinks,citecolor=myBlue,linkcolor=magenta,urlcolor=myBlue]{hyperref} 
\usepackage[capitalise]{cleveref}
\usepackage{tikz}
\usepackage{ulem}

\definecolor{nblue}{rgb}{0.3,0.3,1.0}
\definecolor{ngreen}{rgb}{0.2,0.7,0.2}
\definecolor{nred}{rgb}{0,0,0}
\definecolor{nblack}{rgb}{0,0,0}

 \definecolor{myBlue}{RGB}{3,3,133} 

\begin{document}

\title{Task-Oriented Gaussian Optimization for Non-Gaussian Resources in Continuous-Variable Quantum Computation}

\author{Boxuan~Jing}
\address{State Key Laboratory for Mesoscopic Physics, School of Physics, Frontiers Science Center for Nano-optoelectronics, $\&$ Collaborative Innovation Center of Quantum Matter, Peking University, Beijing 100871, China}
\author{Feng-Xiao~Sun}
\email{sunfengxiao@pku.edu.cn}
\address{State Key Laboratory for Mesoscopic Physics, School of Physics, Frontiers Science Center for Nano-optoelectronics, $\&$ Collaborative Innovation Center of Quantum Matter, Peking University, Beijing 100871, China}
\author{Qiongyi~He}
\address{State Key Laboratory for Mesoscopic Physics, School of Physics, Frontiers Science Center for Nano-optoelectronics, $\&$ Collaborative Innovation Center of Quantum Matter, Peking University, Beijing 100871, China}
\address{Collaborative Innovation Center of Extreme Optics, Shanxi University, Taiyuan, Shanxi 030006, China}
\address{Peking University Yangtze Delta Institute of Optoelectronics, Nantong 226010, Jiangsu, China}		
\address{Hefei National Laboratory, Hefei 230088, China}

\begin{abstract}
In continuous-variable systems, non-Gaussian resources are essential for achieving universal quantum computation that lies beyond classical simulation.
Among the candidate states, the cubic phase state stands out as the simplest form of single-mode non-Gaussian resource, yet its experimental preparation still remains a great challenge.
Although a variety of approximate schemes have been proposed to simulate the cubic phase state, they often fall short when deployed in concrete quantum tasks.
In this work, we present a Gaussian optimization protocol that systematically refines the non-Gaussian resources, which significantly improves the performance of both magic-state-based and measurement-based quantum computation. 
Leveraging task-specific Gaussian operations on approximate cubic phase states, our protocol offers an experimentally feasible approach to enhance gate fidelity in magic-state-based quantum computation and reduce the variance of nonlinear quadrature measurement in measurement-based quantum computation.  
Building on this framework, we further propose a task-oriented non-Gaussian state preparation scheme based on superpositions in the Fock basis followed by squeezing and displacement. 
This strategy enables direct tailoring of resource states to specific task goals. Owing to its flexibility and generality, our framework provides a powerful and broadly applicable tool for enhancing performance across a wide range of continuous-variable quantum information protocols.
\end{abstract}

\maketitle

\section{Introduction}

Quantum computation~\cite{divincenzo1995quantum, nielsen2010quantum} promises to outperform classical information processing in tasks ranging from simulating many-body systems~\cite{ayral2023quantum} to solving classically intractable problems in cryptography~\cite{Shor1999}, optimization~\cite{PRXQuantum.1.020312, PhysRevLett.79.4709,montanaro2016quantum}, and machine learning~\cite{Wang_2024,Schuld03042015}.
Although most experimental progress has been achieved in discrete-variable (DV) platforms such as superconducting qubits \cite{barends2014superconducting,PhysRevLett.107.240501, neill2021accurately} and trapped ions \cite{cirac2000scalable,haffner2008quantum}, continuous-variable (CV) architectures offer an alternative route with scalable resources by encoding quantum information in the infinite-dimensional Hilbert space of bosonic modes \cite{RevModPhys.77.513,RevModPhys.84.621,pfister2019continuous,jia2025continuous,Wang2025Large}.
CV platforms combine deterministic Gaussian operations, flexible state preparation, and compatibility with existing photonic and microwave technologies, making them promising candidates for scalable, fault-tolerant quantum computation \cite{PhysRevLett.112.120504, zhang2024continuous,matsuura2024continuous,PhysRevX.8.021054,PRXQuantum.2.030325,PhysRevLett.130.090602,Niset09errorcor,Lloyd99quantumcomp}.

In CV quantum computing tasks, Gaussian operations play a central role due to their experimental accessibility. However, genuine non-Gaussian resources are indispensable, since Gaussian operations alone cannot implement universal quantum computation \cite{Lloyd99quantumcomp,Menicucci06clusterquantumcomp} or reach quantum advantages \cite{PhysRevLett.109.230503,veitch2012negative}. 
In order to achieve universal CV quantum computation, there have been two main approaches by using cubic phase states. 
The first approach is analogous to the magic state paradigm in DV systems \cite{PhysRevA.71.022316,PhysRevA.86.052329}. 
By combining with Gaussian measurements and feedforward operations, the cubic phase state enables the realization of non-Gaussian gates on arbitrary input states. 
This framework is referred to as magic-state-based quantum computation (MSBQC) \cite{PhysRevResearch.6.023332}.
The second approach exploits the cubic phase state to enable non-Gaussian measurements, which can then be applied in the measurement-based quantum computation (MBQC) \cite{Menicucci06clusterquantumcomp, PhysRevLett.112.120504}.

However, the preparation of ideal cubic phase gates or infinitely squeezed zero-momentum states is still unattainable in practical experiments, resulting in various approximate schemes for finitely squeezed cubic phase states. Examples include schemes through the two-mode squeezing state with Z gate and photon number measurement method \cite{PhysRevA.64.012310}, through the Fock basis superposition method \cite{PhysRevA.88.053816,PhysRevApplied.15.024024,sakaguchi2023nonlinear}, through the photon subtraction method \cite{PhysRevA.95.052352,PhysRevA.91.032321}, through the SNAP gate method \cite{PhysRevLett.115.137002,PRXQuantum.3.030301}, and through the three-photon parametric down-conversion method \cite{PhysRevX.10.011011,PRXQuantum.2.010327,eriksson2024universal}, etc. These approximate schemes generally make trade-offs in terms of fidelity, success rate, and experimental complexity.

Recent advances in Gaussian operation techniques have provided new opportunities for optimizing these approximation schemes. 
Gaussian protocols, characterized by linear optics and squeezing, are intrinsically robust to noise and compatible with existing CV architectures \cite{Walk:16,vandre2025graphical,PRXQuantum.2.030343}.
At present, substantial progress has been made in the control of non-Gaussian states with Gaussian optimization toward higher state fidelity \cite{PhysRevA.105.062446,PRXQuantum.2.010327}. 
However, fidelity is just a measure of the distance between two states in Hilbert space, which cannot reflect the ability of quantum states to perform specific tasks.
Therefore, we hope to directly optimize with Gaussian operations the performance of the approximate cubic phase states in specific quantum computation tasks, which has not yet been explored. 

To address this fundamental challenge, we systematically incorporate Gaussian optimization into three major frameworks to simulate the cubic phase state. 
We propose a unified optimization strategy based on Gaussian operations that substantially enhances their performance in the tasks of both MSBQC and MBQC.
Moreover, to further improve the performance, we carry out a thorough optimization covering the entire process from the preparation of non-Gaussian states to the subsequent Gaussian optimization.
Specifically, we design a scheme that constructs photon-number superposition states followed by squeezing and displacement, with the coefficients optimized via a genetic algorithm. This approach yields superior performance in tasks requiring high similarity to target non-Gaussian states, demonstrating the versatility and effectiveness of our framework.

This paper is structured as follows: Section II reviews the mathematical tools, symbolic conventions, and typical cubic phase state approximation schemes used in this work. Section III presents our Gaussian optimization protocol tailored for enhancing MSBQC. Section IV introduces the application of our protocol to improving MBQC. Section V describes the construction of task-specific non-Gaussian states via Fock basis superpositions optimized using genetic algorithms, and compares their performance with the aforementioned methods. Section VI discusses the experimental feasibility of implementing the optimized non-Gaussian states and associated Gaussian operations. Section VII concludes with a summary and an outlook on future directions.

\section{Cubic phase states}

\begin{figure}
    \centering
    \includegraphics[width=0.45\textwidth]{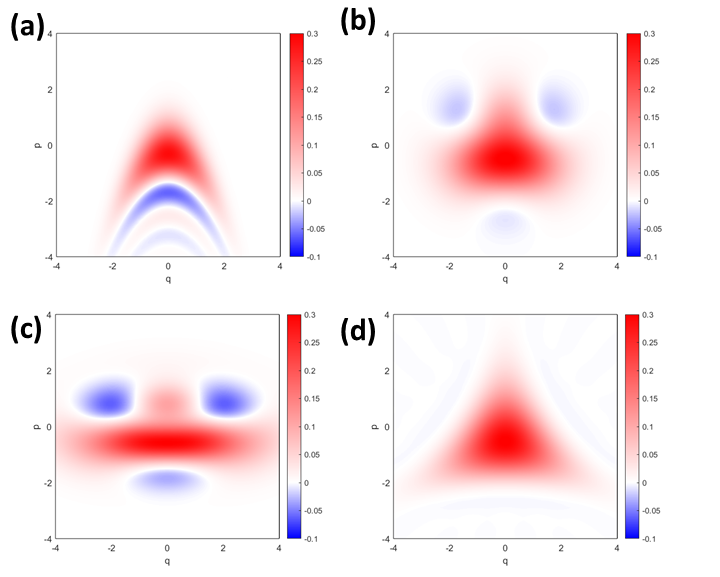}
    \caption{The Wigner functions of (a) finite squeezing cubic phase state $\ket{\phi_c}$, (b) Fock truncation state $\ket{\phi_1}$, (c) superposition squeezing state $\ket{\phi_2}$, (d) three-photon parametric down-conversion optimization state $\ket{\phi_3}$, where $a$=0.173 and $r$=0.3.
    }
\label{fig0}
\end{figure}

An ideal cubic phase state is $e^{ia\hat{q}^3}\ket{0}_p$, where $a$ is the cubic phase strength, $\ket{0}_p$ is the zero-momentum eigenstate, $\hat{q}=(\hat{a}+\hat{a}^\dagger)/{\sqrt{2}}$ and $\hat{p}=i (\hat{a}^\dagger-\hat{a})/{\sqrt{2}}$ are the quadrature operators with $\hat{a}$ being the annihilation operator. In particular, the momentum eigenstate $\ket{p}$ is in the limit of infinite squeezing with normalization $\bra{p'}\ket{p}=\delta_{p,p'}$.
As a result, the zero-momentum state cannot be experimentally realized, and it has a fidelity of zero with practical quantum states. Thus, in general experiments, the finite squeezing cubic phase state is commonly used as a substitute~\cite{PhysRevA.79.062318}, which is expressed as
\begin{align}
\ket{\phi_c}=e^{ia\hat{q}^3}S(-r)\ket{0}.
\end{align}
Here, $S(r)=\exp[(r\hat{a}^2-r\hat{a}^{\dagger^2})/2]$ is the squeezing operator with the squeezing parameter being $r$.
The Wigner function of $\ket{\phi_c}$ has been shown in \cref{fig0}(a). It is obvious that the cubic phase state $\ket{\phi_c}$ is non-Gaussian with negative Wigner distributions.

In order to characterize the non-Gaussianity of quantum states, the method of stellar rank has been developed in recent years~\cite{PhysRevLett.124.063605,PhysRevResearch.3.033018,Chabaud2022holomorphic}. 
Here, the stellar function of a state is defined as $F_\phi(z)=e^{1/2||z||^2}\langle z^*|\phi\rangle$, where $\ket{z}$ is a coherent state of complex amplitude $z$, and the stellar rank is the number of zero points of the stellar function.
A quantum state with a stellar rank of $N$ means that it can be obtained by truncating the initial state with a maximum number of photons of $N$ in the Fock basis and evolving it with Gaussian operations. Notably, a higher stellar rank implies greater difficulty for classical simulation, and thus exhibits more quantum advances~\cite{PhysRevLett.130.090602}.
Obviously, it is directly checked that $\ket{\phi_c}$ is a quantum state with a stellar rank of infinity.

To prepare non-Gaussian states that resemble the finite squeezing cubic phase states, we mainly focus on three approximate schemes in two major categories. One of the categories is to introduce non-Gaussianity under Fock basis with truncation methods, which can be further divided into two schemes: truncating the state in Fock basis (I) or truncating the operator with a weak cubic phase strength (II). The other category is to directly achieve high-order nonlinear interactions, e.g., three-photon parametric down-conversion (III).

For the first approximate scheme (I), we consider truncating the cubic phase state to the Fock basis of $N$=3,
\begin{align}
\ket{\phi_1}=C(c_0\ket{0}+c_1\ket{1}+c_2\ket{2}+c_3\ket{3}),
\end{align}
 where $c_n=\langle n|\phi_c\rangle$ and $C$ is the normalization coefficient. The Wigner function of $\ket{\phi_1}$ can be seen in \cref{fig0}(b).
 Here, $N$=3 is chosen to compare with the following scheme by truncating the operator (II).

\begin{figure}
    \centering
    \includegraphics[width=0.4\textwidth]{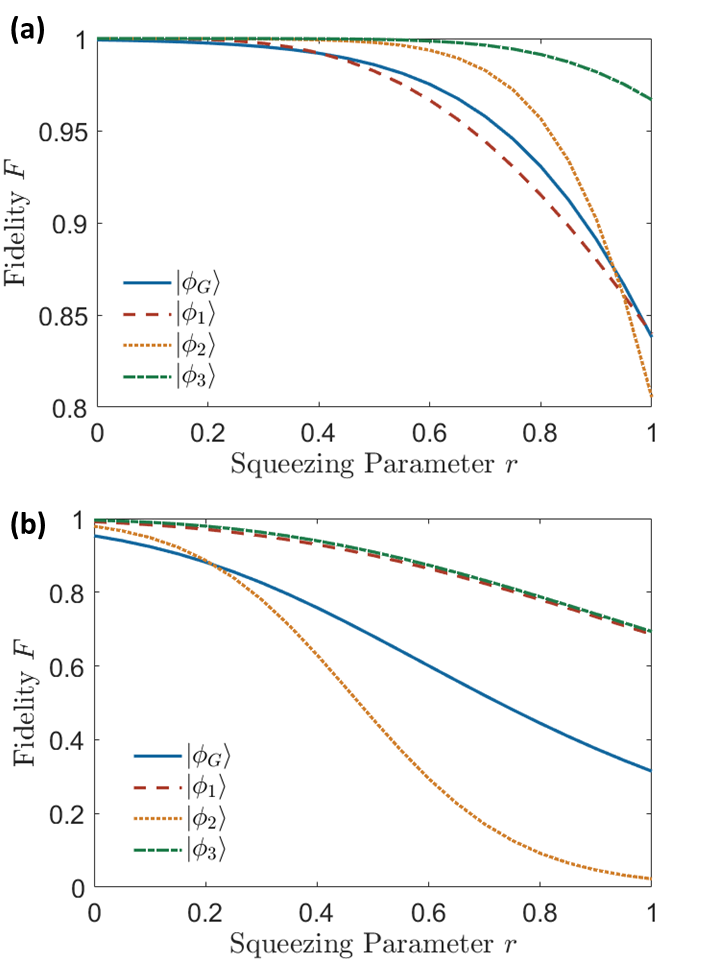}
    \caption{The fidelity comparison of each approximate scheme $\ket{\phi_i}$ and the finite squeezing cubic phase state $\ket{\phi_c}$, where (a) shows $a$=0.02 and (b) shows $a$=0.173.
    }
\label{fig1}
\end{figure}

In the second approximate scheme (II), the cubic phase operator $e^{ia\hat{q}^3}$ is expanded by the Taylor series and truncated so that the state takes the form of
\begin{align}
\ket{\phi_2}=(1+ia\hat{q}^3)S(-r)\ket{0}.\label{eq:phi2}
\end{align}
There are two methods to implement such a truncation experimentally. One method is to factorize the first term $1+ia\hat{q}^3$. As long as the operation of $1+c\hat{q}$ is achieved for any $c$ and performed three times, the non-Gaussian term $1+ia\hat{q}^3$ can be obtained, which has been implemented by photon number projection~\cite{PhysRevA.95.052352}. The other method is to simplify \cref{eq:phi2} as $\ket{\phi'_2}=S(-r)(1+ia\hat{q}^3e^{3r})\ket{0}$. It is worth noting that the item $(1+ia\hat{q}^3e^{3r})\ket{0}$ represents a photon superposition state where the number of photons does not exceed 3. Although these two methods seem different, the quantum states achieved are actually equivalent. The Wigner function of $\ket{\phi_2}$ is shown in \cref{fig0}(c). And its stellar rank is 3, the same as $\ket{\phi_1}$.

The third approximate scheme (III) that directly utilizes high-order nonlinear effects benefits from the recent development of the experimental technology of three-photon parametric down-conversion ~\cite{PhysRevX.10.011011}.
It has shown that by adding displacement and squeezing on the basis of the three-photon parametric down-conversion state, the obtained state of
\begin{align}
\ket{\phi_3}=D_p(s)S(t)e^{if\hat{a}^3+if^*\hat{a}^{\dagger^3}}\ket{0}
\end{align}
can simulate the finite squeezing cubic phase state $\ket{\phi_c}$ with a high fidelity \cite{PRXQuantum.2.010327}. The stellar rank of $\ket{\phi_3}$ is found to be infinity, whose Wigner function is presented in \cref{fig0}(d).

It is evident from \cref{fig0} that the Wigner distributions of the quantum states obtained in approximate schemes differ markedly from that of the finite squeezing cubic phase state $\ket{\phi_c}$. Thus, fidelity is commonly used to evaluate these schemes. In \cref{fig1}, we show the fidelity of each approximate scheme and the finite squeezing cubic phase state.
The detailed calculations can be found in \cref{app:1}.
In addition, in order to demonstrate the advantages of non-Gaussianity, we take the pure Gaussian squeezed state as the benchmark for comparison, which takes the form of 
\begin{align}
\ket{\phi_G}=S(-r)\ket{0}.
\end{align}
It can be considered that this state is the result of expanding the cubic phase gate in front of $\ket{\phi_c}$ to the 0th order term, and its stellar rank is 0.

As shown in Fig.~\ref{fig1}(a), when the cubic phase strength $a$ is very weak, all these approximation schemes, including the Gaussian squeezed state $\ket{\phi_G}$, maintain fidelity \(F\geq0.95\) up to moderate squeezing (\(r\leq0.6\)). Among them, the three‐photon down‐conversion protocol \(\lvert\phi_3\rangle\) (green curve) yields the highest fidelity throughout the entire \(r\)-range, while the other three schemes show quick drops when $r$ becomes large. 

In fact, universal quantum computing requires the implementation of cubic phase gates with arbitrary $a$ for the decomposition of CV gates  \cite{PhysRevLett.107.170501,PhysRevResearch.6.023332}. 
Therefore, we further examine these approximation schemes under the condition of large $a$ in \cref{fig1}(b), which shows that all the state fidelities of these schemes decrease.
Whereas, both the Fock truncation $\ket{\phi_1}$ and the three-photon parametric down-conversion $\ket{\phi_3}$ are significantly better than the other two schemes in terms of fidelity. It is worth noting that the fidelities of these two schemes with the finite squeezing cubic phase state are always very similar, but these two states are actually very different. 
It can be clearly seen from \cref{fig0} that the Wigner negativity of $\ket{\phi_3}$ is significantly less than that of $\ket{\phi_1}$.
This indicates that for larger nonlinearity, a finite photon number truncation offers the most robust route to simulating the ideal squeezing cubic phase state.

\section{Gaussian optimization for magic-state-based quantum computation}

\begin{figure}
    \centering
    \includegraphics[width=0.5\textwidth]{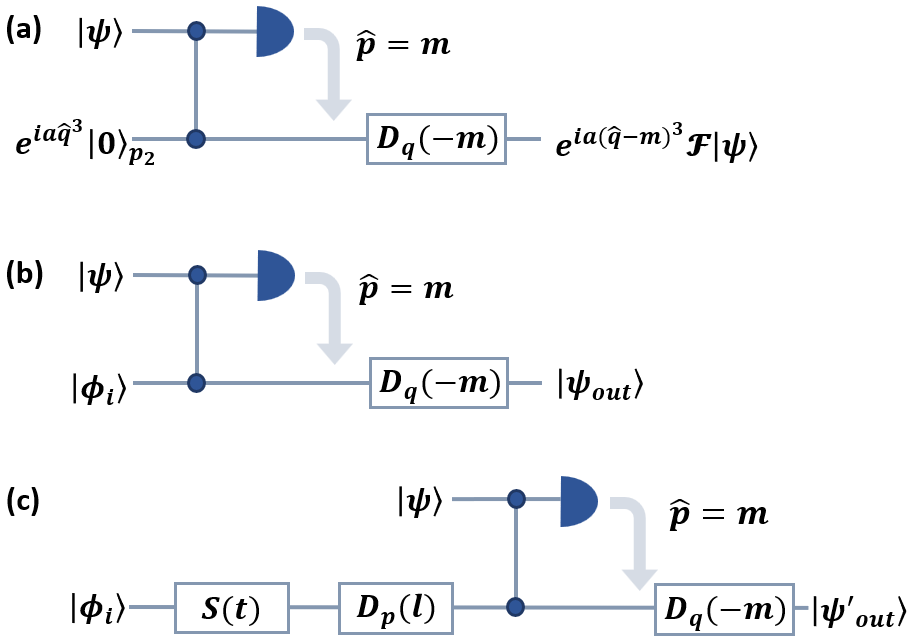}
    \caption{
    (a) A simple quantum circuit model for MSBQC, which associates the cubic phase state and arbitrary initial state $\ket{\psi}$ through a controlled-Z gate, then performs a $\hat{p}$ measurement on the $\ket{\psi}$ mode to obtain $m$, and performs a $D_q(-m)$ displacement operation on the cubic phase state. The output state can be calculated by applying a displacement cubic phase gate to the Fourier transform of $\ket{\psi}$.
    (b) Replace the cubic phase state with $\ket{\phi_i}$ prepared by an approximate scheme to obtain the output state $\ket{\psi_{out}}$.
    (c) Add our Gaussian optimization operations including squeezing and momentum displacement to obtain the output state  $\ket{\psi'_{out}}$.
    }
\label{fig2diagram}
\end{figure}

Here we focus on the MSBQC task, which is described by the circuit shown in Fig.~\ref{fig2diagram}(a). When $m=0$, this circuit achieves the effect of applying the cubic phase gate to the initial state $\ket{\psi}$.
If there is an ideal cubic phase state and an arbitrary initial state $\ket{\psi}=\int_{-\infty}^\infty f(q)\ket{q}dq$, by connecting them through a controlled-Z gate and then performing the $\hat{p}$ measurement on the $\ket{\psi}$ mode, we can get the conditional state,
\begin{align}
\nonumber
\ket{\phi_{temp}}
&= e^{i a q_2^3}
   \int_{-\infty}^\infty \bra{p_1=m}f(q)\ket{q}_{q_1}\,\ket{q}_{p_2}\,\mathrm{d}q
\\
&= e^{i a q_2^3}
   \int_{-\infty}^\infty e^{i m q} \,f(q)\,\ket{q}_{p_2}\,\mathrm{d}q.
\end{align}
Then we perform a displacement $D_q(-m)$ based on the measurement outcome $m$ so that the output state $e^{ia(\hat{q}-m)^3}\mathcal{F}\ket{\psi}$ is obtained, where $D_q(s_q)=e^{is_q\hat{p}}$ [$D_p(s_p)=e^{is_p\hat{q}}$] is the position (momentum) displacement operator, and $\mathcal{F}$ represents the Fourier transform that satisfies $\mathcal{F}\ket{s}_q=\ket{s}_p$.
In this way, the cubic phase gate can be applied to arbitrary quantum states, and theoretically it can be extended to general quantum computing tasks.

\begin{figure}
    \centering
    \includegraphics[width=0.43\textwidth]{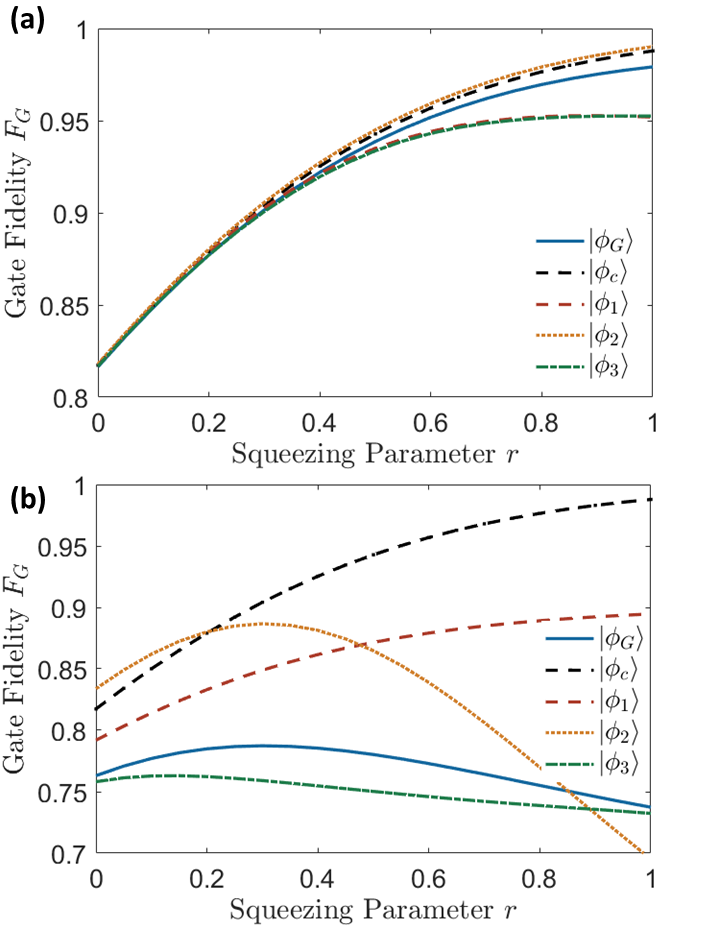}
    \caption{The gate fidelity, which is obtained by the fidelity of the output state without optimization $\ket{\psi_{out}}$ and the ideal output state $e^{ia(\hat{q}-m)^3}\mathcal{F}\ket{\psi}$, where (a) shows a=0.02 and (b) shows a=0.173.
    }
\label{fig3a}
\end{figure}

Now we consider the practical experimental conditions and replace the ideal cubic phase with a certain approximate preparation scheme $\ket{\phi_i}$, as shown in Fig.~\ref{fig2diagram}(b). 
We aim to evaluate the fidelity of this quantum computational circuit gate.
Assume that the measurement result is $m=0$, and the initial input state $\ket{\psi}$ is a squeezed state $S(R)\ket{0}$ with $R=0.5$.
In general, the fidelity of the cubic phase gate is defined as the statistical average of the overlaps between the ideal and the actual output states over a distribution of input states. For convenience, in this work we replace this averaging process by evaluating the fidelity with respect to a fixed squeezed state. That is, the gate fidelity $F_G$ is given by
\begin{align}
F_G=|\langle \psi_{out} | e^{ia(\hat{q}-m)^3}\mathcal{F}\ket{\psi}|^2.
\end{align}
The ideal computational output state $e^{ia(\hat{q}-m)^3}\mathcal{F}\ket{\psi}$ is illustrated in \cref{fig2diagram}(a) and the actual output state $\ket{\psi_{out}}$ is depicted in \cref{fig2diagram}(b).
The detailed calculations can be found in \cref{app:2}, and the gate fidelity is demonstrated in \cref{fig3a}.

It can be found from \cref{fig3a} that all the approximate preparation schemes can achieve relatively high-fidelity quantum computing. 
Surprisingly, the three-photon parametric down-conversion state $\ket{\phi_3}$, which has the highest fidelity in \cref{fig1}, exhibits the worst performance in quantum computing tasks (even worse than the Gaussian state $\ket{\phi_G}$).
This indicates that fidelity of states is not a good indicator for the execution of specific tasks.
Meanwhile, the truncation method $\ket{\phi_1}$ is still very stable when $a$ is relatively large, while the superposition squeezing method $\ket{\phi_2}$ shows the strongest computing power when the squeezing $r$ and the nonlinear intensity $a$ are low. 
In particular, when $r$ is increased to around $1$ with a large cubic nonlinearity $a$, the fidelity of the state $\ket{\phi_2}$ itself drops to near $0$ [\cref{fig1}(b)], while the gate fidelity still remains about $0.75$ [\cref{fig3a}(b)].
This suggests that different preparation schemes should be suitable for different quantum information tasks. Thus, task-oriented schemes need to be developed.

\begin{figure}
    \centering
    \includegraphics[width=0.43\textwidth]{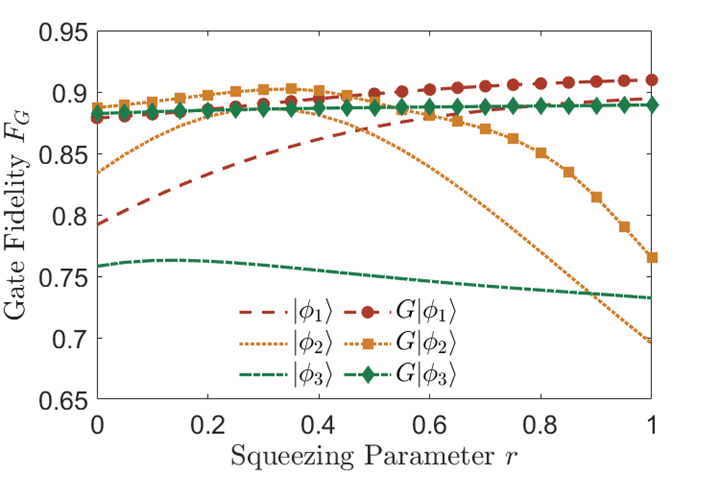}
    \caption{Comparison of gate fidelities with and without Gaussian optimization. $G$ represents the corresponding optimized displacement and squeezing operations for each $\ket{\phi_i}$, with the parameter set to $a$ = 0.173.  
    }
\label{fig4}
\end{figure}

In order to address challenges in quantum computing arising from imperfections in experimental quantum state preparation, we involve a Gaussian optimization protocol in the circuit as shown in Fig. \ref{fig2diagram}(c).
It has been proven that any single-mode Gaussian operation can be decomposed into displacement, squeezing, and rotation~\cite{Weedbrook_2012_05}. 
Considering that the quantum states $\ket{\phi_i}$ possess reflection symmetry about the $q=0$ axis in phase space, we exclude position displacement and rotation, and instead consider only momentum displacement and squeezing transformations, to preserve this structural property during optimization.
In the following, we will show that this Gaussian optimization can enhance the gate fidelity in MSBQC tasks.

In Fig. \ref{fig4}, it is revealed that when this Gaussian optimization is applied, the differences in gate fidelity among these approximate schemes become small, and their performance in MSBQC is substantially improved.
Specifically, the gate fidelity of $\ket{\phi_1}$ and $\ket{\phi_3}$ is very stable around $0.9$ after optimization. 

\section{Gaussian optimization for measurement-based quantum computation}

\begin{figure}
    \centering
    \includegraphics[width=0.52\textwidth]{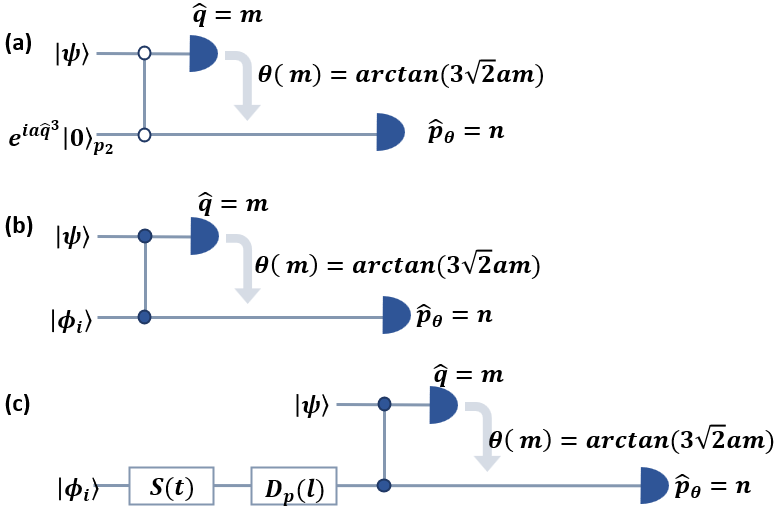}
    \caption{
     (a) A simple quantum circuit for MBQC, which associates the cubic phase state and an arbitrary initial state $\ket{\psi}$ through a beam splitter, then performs $\hat{q}$ measurement on the $\ket{\psi}$ mode to obtain $m$.
     After a nonlinear data processing, the measurement is conducted at $\hat{p}_\theta=\hat{p}cos\theta+\hat{q}sin\theta$ on the cubic phase state and the result is $n$.
     Then we can achieve the nonlinear measurement of $\hat{p}_{in}-3a\hat{q}_{in}^2=\sqrt{2}n/cos\theta$.
     (b) Replace the cubic phase state with $\ket{\phi_i}$ prepared by an approximate scheme. Then the nonlinear measurement of $\hat{p}_{in}-3a\hat{q}_{in}^2+\hat{p}_{\phi_i}-3a\hat{q}_{\phi_i}^2=\sqrt{2}n/cos\theta$ is obtained. 
     (c) Add our Gaussian optimization including squeezing and momentum displacement on $\ket{\phi_i}$ to minimize the variance of the nonlinear measurement.
    }
\label{fig5}
\end{figure}

Due to the difficulty in preparing quantum gates in quantum computing tasks, the MBQC theory was developed, which aims to replace non-Gaussian gates with nonlinear quadrature measurements \cite{Menicucci06clusterquantumcomp, PhysRevLett.112.120504}. In previous research \cite{sakaguchi2023nonlinear}, cubic phase states have been proposed to realize nonlinear measurements, as shown in Fig.~\ref{fig5}(a). For an arbitrary input state $\ket{\psi}$, it is associated with the cubic phase state using a beam splitter. Then $\hat{q}$ is measured in the $\ket{\psi}$ mode, and $\hat{p}_\theta$ is measured in the cubic phase state mode. Finally, the measurement results are post-processed to obtain the nonlinear measurement results of $\ket{\psi}$. This measurement circuit can achieve equivalent nonlinear quadrature measurement results of $\hat{p}-3a\hat{q}^2$ through linear measurement, in which the error of nonlinear measurement is directly related to the accuracy of the cubic phase state. Therefore, we hope to improve the accuracy of nonlinear measurement by optimizing the approximate cubic phase states. In this work, we use the variance of $\hat{p}-3a\hat{q}^2$ to quantify the accuracy, where a lower variance represents a better performance of the MBQC task.

\begin{figure}
    \centering
    \includegraphics[width=0.43\textwidth]{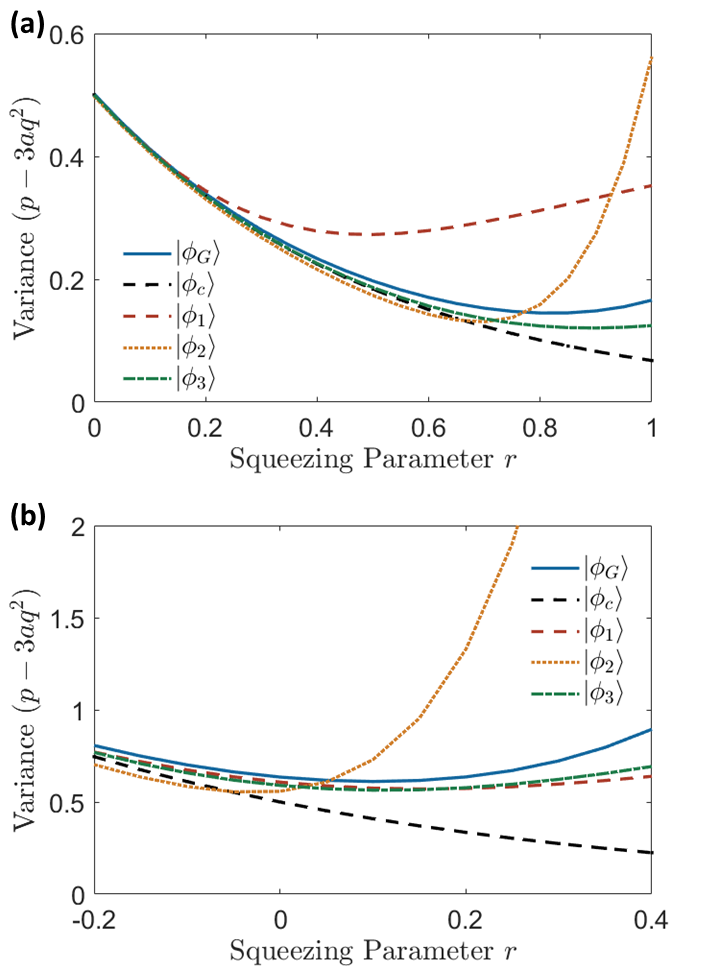}
    \caption{The variances of nonlinear measurement $\hat{p}-3a\hat{q}^2$ for the approximate cubic phase states and the Gaussian squeezing state $\ket{\phi_G}$, where (a) a=0.02 and (b) a=0.173.
    }
\label{fig6}
\end{figure}

In practical experiments, the ideal cubic phase state has to be achieved by approximate schemes. Thus, the MBQC circuit is modified as in Fig.~\ref{fig5}(b). With the approximate schemes mentioned, we show the relationship between the variance and the squeeze parameter $r$ in Fig.~\ref{fig6} under the conditions of $a=0.02$ and $a=0.173$. The detailed calculations can be found in \cref{app:3}. 

It is found in Fig.~\ref{fig6}(a) that when $a$ is relatively small, all the mentioned preparation schemes can provide nonlinear measurement capabilities that exceed the shot noise limit. But when the squeezing parameter $r$ is relatively large, there will be a significant deviation for these approximate schemes. It is also observed that in such a case, both $\ket{\phi_1}$ and $\ket{\phi_2}$ exhibit a worse performance than the Gaussian squeezed state $\ket{\phi_G}$.

Then we turn to Fig.~\ref{fig6}(b) and verified that the Gaussian state response of the variance is $0.64$ at the point of $r=0$ for $\ket{\phi_G}$ and $0.56$ at the point of $r=0$ for $\ket{\phi_2}$, which is consistent with the theoretical and experimental results in Ref.~\cite{sakaguchi2023nonlinear}. It is also found when the cubic phase strength $a$ and the squeezing parameter $r$ become large, all the variances increase exponentially with $r$ except for the finite squeezing cubic phase $\ket{\phi_c}$. This means that it is difficult to obtain larger nonlinear measurement accuracies without optimization.

\begin{figure}
    \centering
    \includegraphics[width=0.43\textwidth]{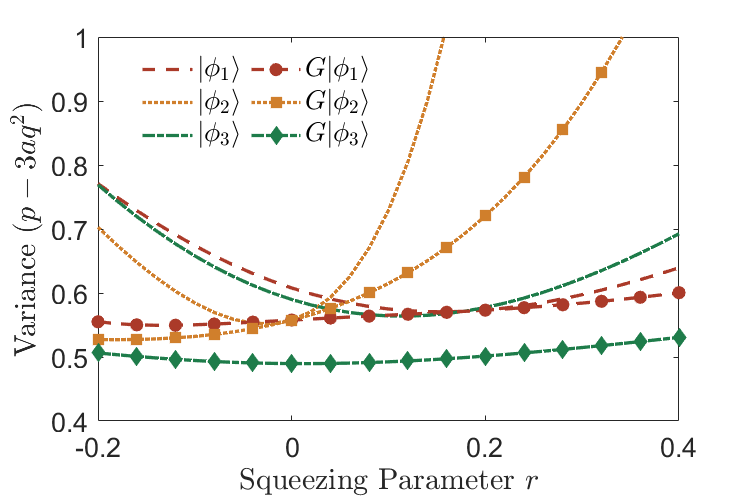}
    \caption{
 Comparison of the variance of $\hat{p}-3a\hat{q}^2$ of the approximate cubic phase states $\ket{\phi_i}$ and the optimized $G\ket{\phi_i}$.
 $G$ represents the corresponding optimized displacement and squeezing operation, with the parameter set to $a$ = 0.173.
    }
\label{fig7}
\end{figure}

Then we use Gaussian operations to minimize the variance of the nonlinear measurement $\hat{p}-3a\hat{q}^2$ for approximate cubic phase states $\ket{\phi_1}$, $\ket{\phi_2}$, and $\ket{\phi_3}$ with the circuit shown in Fig.~\ref{fig5}(c), where the results are presented in Fig.~\ref{fig7}. It can be clearly seen that the variance after optimization is smaller than the variance before optimization. For the preparation scheme $\ket{\phi_2}$, the nonlinear measurement variance can be optimized to $0.52$ through Gaussian operations. And for $\ket{\phi_3}$, whose performance of the optimized gate fidelity in MSBQC is not the best, the nonlinear measurement accuracy behaves the best among these approximate schemes for MBQC, where the optimized variance reaches $0.48$. This indicates that different approximate schemes are suitable for different quantum computing tasks. In other words, the use of task-oriented schemes can enhance the operational efficiency of each specific task. Furthermore, the Gaussian optimization protocol we proposed can always be applied to improve non-Gaussian resources for quantum information processing, regardless of the specific task or scheme.

\section{A general task-oriented non-Gaussian state preparation scheme for global optimization}

\begin{table*}[t]
\caption{When the maximum number of photons in the initial Fock basis superposition is $N$, the coefficients of the Fock basis use the angles $\theta_i$ and $\phi_j$ of the high-dimensional Bloch sphere as parameters, with a squeezing gate $S(r)$ and a displacement gate $D_q(q)$ applied. The parameter results are optimized with the gate fidelity as the optimization goal when $a=0.173$.}
\begin{ruledtabular}
\begin{tabular}{lccccccccccccc}
$N$ & $q$ & $r$ & $\theta_1$ & $\phi_1$ & $\theta_2$ & $\phi_2$ & $\theta_3$ & $\phi_3$ & $\theta_4$ & $\phi_4$ & $\theta_5$ & $\phi_5$ &  Fid\\
\hline
0 & 0.2924 & -0.672 &  &  &  &  &  &  &  &  &  &  & 0.871\\
1 & 0.871 & -0.246 & 0.904 & 4.712 &  &  &  &  &  &  &  &  & 0.938\\
2 & 1.369 & -0.063 & 1.411 & 4.712 & 0.7131 & 3.14 &  &  &  &  &  &  & 0.961\\
3 & 1.278 & -0.054 & 1.328 & 4.712 & 0.588 & 3.14 & 0.192 & 4.712 &  &  &  &  & 0.961\\
4 & 1.0835 & -0.31 & 1.196 & 4.712 & 0.5115 & 3.14 & 1.57 & 4.712 & 1.555 & 0 &  &  & 0.979\\
5 & 1.4655 & -0.1737 & 1.549 & 4.712 & 0.8929 & 3.14 & 0.5605 & 4.712 & 1.283 & 0 & 1.299 & 4.712 & 0.987\\
\end{tabular}
\end{ruledtabular}
\label{table1}
\end{table*}

\begin{table*}[t]
\caption{When the maximum number of photons in the initial Fock basis superposition is $N$, the coefficients of the Fock basis use the angles $\theta_i$ and $\phi_j$ of the high-dimensional Bloch sphere as parameters, with a squeezing gate $S(r)$ and a displacement gate $D_q(q)$ applied. The parameter results are minimized with the variance of the nonlinear quadrature measurement when $a=0.173$.}
\begin{ruledtabular}
\begin{tabular}{lccccccccccccc}
$N$ & $q$ & $r$ & $\theta_1$ & $\phi_1$ & $\theta_2$ & $\phi_2$ & $\theta_3$ & $\phi_3$ & $\theta_4$ & $\phi_4$ & $\theta_5$ & $\phi_5$ &  Var\\
\hline
0 & 0 & -0.1025 &  &  &  &  &  &  &  &  &  &  & 0.611\\
1 & 0 & 0.02 & 0.6602 & 4.712 &  &  &  &  &  &  &  &  & 0.438\\
2 & 0 & 0.1015 & 1.12 & 4.712 & 0.5197 & 3.14 &  &  &  &  &  &  & 0.361\\
3 & 0 & 0.1625 & 1.403 & 4.712 & 0.9658 & 3.14 & 0.4502 & 1.57 &  &  &  &  & 0.316\\
4 & 0 & -0.0407 & 0.9835 & 4.712 & 0.3156 & 3.14 & 0.5511 & 1.57 & 1.57 & 0 &  &  & 0.308\\
5 & 0 & 0.037 & 1.323 & 4.712 & 0.7534 & 3.14 & 0.2648 & 1.57 & 0.8175 & 0 & 1.25 & 4.713 & 0.265\\
\end{tabular}
\end{ruledtabular}
\label{table2}
\end{table*}

As demonstrated in the previous sections, existing approximation schemes for preparing cubic phase states tend to exhibit growing deviations from the ideal cubic phase state when the nonlinearity strength \( a \) increases. 
Moreover, contrary to intuitive expectations, increasing the squeezing strength does not consistently improve performance in quantum computing tasks. 
Although Gaussian optimization via squeezing and displacement can enhance gate fidelity in MSBQC and nonlinear measurement accuracy in MBQC, each approximation strategy exhibits intrinsic limits beyond which no further improvement is achievable.

To overcome these limitations, we propose a more general non-Gaussian state preparation framework based on the concept of stellar rank. Specifically, we construct a state of the form  
\begin{align}
\ket{\phi_4} = D(q)\,S(r) \sum_{i=0}^{N} c_i \ket{i},
\end{align}
where a finite Fock basis superposition is followed by squeezing and displacement operations. The coefficients \( \{c_i\} \) are optimized to further improve the task-specific performance. 
We use the parameterization method on the high-dimensional complex Bloch sphere,
\begin{equation}
c_{j} = 
\begin{cases}
\cos\theta_1, & j = 0, \\
\left( \prod\limits_{k=1}^{j} \sin\theta_k \right) \cos\theta_{j+1} \, e^{i\phi_j}, & 1 \le j \le N-1, \\
\prod\limits_{k=1}^N \sin\theta_k \, e^{i\phi_N}, & j = N.
\end{cases}
\end{equation}

To efficiently search the high-dimensional parameter space, we employ a genetic algorithm to optimize the photon-number-dependent superposition states.
Genetic algorithms are population-based stochastic optimization methods inspired by the principles of natural selection and evolution in biology \cite{genetic,mitchell}.
In our implementation, a population of candidate solutions $\{\theta_i,\phi_j,r,q \}$ is initialized randomly, where $\{\theta_i,\phi_j \}$ determine the Bloch-sphere parameterization of the superposition coefficients, and $\{r,q \}$ represent the squeezing and displacement operations, respectively.
Each candidate state is then evaluated according to a task-specific fitness function, such as the gate fidelity or the variance of $\hat{p}-3a\hat{q}^2$.
The best-performing candidates are selected to form the next generation through reproduction operations, which include crossover (recombination of parameter sets from two parents) and mutation (random perturbations to avoid local minima).
This iterative process is repeated until the convergence criteria are met.

This evolutionary approach is especially well-suited to our problem, as the optimization landscape of non-Gaussian state preparation is typically non-convex with numerous local optima, making gradient-based methods less effective.
In contrast, genetic algorithms provide a global search capability and demonstrate robustness in high-dimensional parameter spaces.

As a result, we present in Tables~\ref{table1} and~\ref{table2} the optimized parameters and resulting performance under two distinct objectives: maximizing the gate fidelity for MSBQC and minimizing the variance of nonlinear quadrature measurement for MBQC, respectively. The optimization is performed with photon number truncation \( N < 6 \). 

As shown in Fig. \ref{fig5}, existing preparation schemes can only achieve a gate fidelity of approximately 0.9 in MSBQC, even after Gaussian optimization. In contrast, under our global optimization, a superposition state with only $N=1$ photon already reaches a gate fidelity of 0.938. Moreover, the fidelity can steadily increase as the number of superimposed photons grows.

For MBQC tasks, existing schemes achieve only a nonlinear variance of about 0.5 after Gaussian optimization as illustrated in Fig. \ref{fig7}. Our global optimization, however, reduces the nonlinear variance to 0.438 with just a single-photon superposition $N=1$. Furthermore, the nonlinear variance can be further suppressed as the number of superimposed photons increases. 
These results demonstrate that high-fidelity quantum computation does not necessarily require structurally complicated non-Gaussian states with high stellar ranks. 
Conversely, the non-Gaussian states in our task-oriented preparation scheme have the potential to offer more effective and resource-efficient performance in quantum computation.

\section{Experimental Feasibility}

The experimental feasibility analysis is divided into two parts, corresponding to the two main components of our proposed protocol. 

The first component enhances existing non-Gaussian state preparation schemes by applying additional Gaussian operations, including squeezing and displacement. This Gaussian-enhanced approach is platform-independent and, in principle, can be applied to optimize any non-Gaussian state for arbitrary tasks \cite{PhysRevA.79.012313}. Its generality and flexibility make it compatible with a wide range of experimental systems \cite{PhysRevA.88.043818,PhysRevA.100.013831,PhysRevA.90.063809}.

High-purity non-Gaussian states have been experimentally realized using photon subtraction and quantum shearing. Deterministic linear squeezing has been achieved using optical parametric amplifiers, which apply \(-3\,\mathrm{dB}\) of squeezing to a single-photon state to produce a Schrödinger cat state with a fidelity of 92\%~\cite{PhysRevLett.113.013601}. Displacement operations are routinely performed using linear optics components such as beam splitters and coherent state injection, with phase-locking accuracy on the order of \(10^{-3}\,\mathrm{rad}\), and have been shown to be fully compatible with squeezing~\cite{PhysRevApplied.16.064037}.

The second component is based on the preparation of arbitrary superpositions of Fock states. 
Ref.~\cite{PhysRevA.72.033822} proposed a method to generate arbitrary superpositions of photon-number states using squeezed states, photon detection, and displacement operations. Based on this approach, superpositions up to three photons have been demonstrated in CV quantum optics platforms~\cite{Yukawa:13}. 

Finally, by comparing the parameters required in our scheme (Tables~\ref{table1} and~\ref{table2}) with the above-mentioned experiments, we find that all components of our protocol are well within the reach of current experimental technologies, making our proposed approach viable for near-term implementation and valuable for high-fidelity quantum information processing.

\section{Conclusion}

We have presented a Gaussian optimization protocol that upgrades existing approximations of the cubic phase state for quantum computation. Applied to three frameworks, our Gaussian optimization protocol enhances gate fidelity for MSBQC and lowers nonlinear measurement variance for MBQC, even when the underlying state fidelity is limited.
In addition, we further design a task-oriented non-Gaussian state preparation and optimization scheme based on photon-number superpositions with a genetic algorithm. Remarkably, superior performance is achieved with only a single-photon component, outperforming current schemes while remaining experimentally accessible.
Beyond the cubic phase states presented here, our approach offers a modular framework for resource-state engineering. Any approximate non-Gaussian state can be directly integrated and optimized, reducing the cost of high-precision operations in CV platforms. This pathway makes scalable high-order gates more experimentally accessible, and may also open new opportunities in quantum sensing, communication, and computation.

\begin{acknowledgments}
This work is supported by the National Natural Science Foundation of China (Grants No. 12125402, No. 12534016 and No. 12474256), the Innovation Program for Quantum Science and Technology (Grants No. 2024ZD0302401 and No. 2021ZD0301500), and Beijing Natural Science Foundation (Grant No. Z240007).
\end{acknowledgments}

\appendix

\section{The expansion of quantum states in the $\hat{q}$ space}\label{app:1}

For the sake of convenience, we calculate the wave functions of these quantum states in a unified form in position space.
\begin{align}
\ket{\phi_G}=S(-r)\ket{0}=\frac{1}{\sqrt{e^r\sqrt{\pi}}}\int e^{-q^2e^{-2r}/2}\ket{q}dq.
\end{align}
\begin{align}
\ket{\phi_c}=e^{iaq^3}S(-r)\ket{0}=\frac{1}{\sqrt{e^r\sqrt{\pi}}}\int e^{iaq^3}e^{-q^2e^{-2r}/2}\ket{q}dq.
\end{align}
\begin{align}
\ket{\phi_1}=C_0\ket{0}+C_1\ket{1}+C_2\ket{2}+C_3\ket{3},
\end{align}
 where $C_n={c_n}/{\sqrt{c_0^2+c_1^2+c_2^2+c_3^2}}$, $c_n=\langle n|\phi_1\rangle=\frac{1}{\sqrt{e^r\pi2^n n!}}\int e^{iaq^3}e^{-q^2(e^{-2r}+1)/2}H_n(q)\ket{q}dq$, and $H_n(q)$ is the Hermite polynomial. 
\begin{align}\nonumber
\ket{\phi_2}
&= (1 + i a \,\hat q^3)\,S(-r)\ket{0} \\
&= \frac{1}{\sqrt{e^r\sqrt{\pi}\bigl(1+\tfrac{15}{8}a^2 e^{6r}\bigr)}} 
   \int\!(1 + i a q^3)\,e^{-q^2 e^{-2r}/2}\,\ket{q}\,\mathrm{d}q.
\end{align}
\begin{align}
\ket{\phi_3}
&= C \int e^{i s q}
   \exp\!\Bigl[\frac{i f}{\sqrt8}
     \bigl((e^{-t}q + e^{t}\partial_q)^3
          + (e^{-t}q - e^{t}\partial_q)^3\bigr)
   \Bigr] \nonumber\\
&\qquad\times
   \exp\!\Bigl(-\tfrac{e^{-2t}q^2}{2}\Bigr)
   \,\ket{q}\,\mathrm{d}q.
\end{align}

Since all quantum states are expanded in the $\hat{q}$ basis, we can naturally obtain the integral expression for all fidelity

\begin{align}
F_i=|\langle\phi_i|\phi_1\rangle|^2=|\int_{-\infty}^\infty\phi_i(q)^*\phi_1(q)dq|^2,
\end{align}
where $\ket{\phi_i}=\int \phi_i(q)\ket{q}dq$.

When we consider Gaussian operation optimization, it is equivalent to adding squeezing and displacement operations before $\ket{\phi_i}$ to make it become $G\ket{\phi_i}=D_p(s)S(t)\ket{\phi_i}=C\int e^{isq}\phi_i(e^tq)\ket{q}dq$.

Hence, for each specific form of the task, we can calculate a final task result based on $\phi_i(q)$. Replacing $\phi_i(q)$ with $\phi_j(q)$ reflects different preparation schemes. And the Gaussian optimization process is calculated by replacing $\phi_i(q)$ with $e^{isq}\phi_i(e^tq)$ and optimizing the parameters of $s$ and $t$.

\section{Results and Optimization of MSBQC}\label{app:2}

Suppose the initial state of any input is $\ket{\psi}=\int \psi(q) \ket{q} dq$ and the auxiliary state is $\ket{\phi_i}=\int \phi_i(q) \ket{q} dq$, then the output state passing through the circuit in \cref{fig3a} is
\begin{align}\nonumber
\ket{\psi_{out}}&=X(-m)\bra{p_1=m}\int e^{iq_1q_2}\phi_i(q_2)\psi(q_1)\ket{q_1q_2}dq_1dq_2\\
&=\int e^{iq_1q_2}\psi(q_1)dq_1\int \phi_i \ket{q_2}dq_2.
\end{align}
Here we consider $m=0$, and $\psi(q)=e^{-q^2e^{-2R}/2}$, so $\ket{\psi_{out}}=e^{-\hat{q}^2e^{2R}/2}\int \phi_i(q)\ket{q}dq$.
For an ideal cubic phase state, we have
\begin{align}
\ket{\psi_{ideal}}=(\frac{e^{2R}}{\pi})^{1/4}\int e^{iaq^3}e^{-q^2 e^{2R}/2}\ket{q}dq.
\end{align}
By the method given in Appendix \ref{app:1}, we can calculate the fidelity of the optimized output states $\ket{\psi_{out}}$ and $\ket{\psi_{ideal}}$
\begin{align}
F_{out_i}=|(\frac{e^{2R}}{\pi})^{1/4}\int_{-\infty}^\infty e^{-iaq^3}e^{-q^2e^{2R}/2}\phi_i^*(q)dq|^2.
\end{align}

\section{Results and Optimization of MBQC}\label{app:3}

To reduce the variance of the nonlinear quadrature measurements, it is necessary to reduce the Var($p-3aq^2$) of the input state $\ket{\phi_i}$. For the nonlinear quadrature measurements of arbitrary state, we have
\begin{align}\nonumber
Var(\hat{p}-3a\hat{q}^2)&=E((\hat{p}-3a\hat{q}^2)^2)-E^2(\hat{p}-3a\hat{q}^2)\\\nonumber
&=E(\hat{p}^2)-3aE(\hat{p}\hat{q}^2)-3aE(\hat{q}^2\hat{p})\\\nonumber
&\hspace{1em}+9a^2E(\hat{q}^4)-E(\hat{p})^2-9a^2E^2(\hat{q}^2)\\
&\hspace{1em}+6aE(\hat{p})E(\hat{q}^2).
\end{align}

Considering $p=-i\frac{d}{dq}$, all the expected values can be simplified,
\begin{align}
E(\hat{p}^2)=\int_{-\infty}^\infty \phi_i^*(q)(-i\frac{d}{dq})^2\phi_i(q)dq,
\end{align}
\begin{align}
E(\hat{p}\hat{q}^2)=\int_{-\infty}^\infty \phi_i^*(q)(-i\frac{d}{dq})q^2\phi_i(q)dq,
\end{align}
\begin{align}
E(\hat{q}^2\hat{p})=\int_{-\infty}^\infty \phi_i^*(q)q^2(-i\frac{d}{dq})\phi_i(q)dq,
\end{align}
\begin{align}
E(\hat{q}^4)=\int_{-\infty}^\infty \phi_i^*(q)q^4\phi_i(q)dq,
\end{align}
\begin{align}
E(\hat{p})=\int_{-\infty}^\infty \phi_i^*(q)(-i\frac{d}{dq})\phi_i(q)dq,
\end{align}
\begin{align}
E(\hat{q}^2)=\int_{-\infty}^\infty \phi_i^*(q)q^2\phi_i(q)dq.
\end{align}
Then we can obtain the variance of the nonlinear quadrature measurements through these integration results for each quantum state.

\bibliography{references}
\end{document}